\documentclass[conference]{IEEEtran}
\IEEEoverridecommandlockouts
\usepackage{cite}
\usepackage{amsmath,amssymb,amsfonts}
\usepackage{algorithmic}
\usepackage{graphicx}
\usepackage{textcomp}
\usepackage{xcolor}
\usepackage{balance}
\usepackage{cite}
\usepackage{hyperref}
\usepackage{amsmath,amssymb,amsfonts}
\usepackage{amsmath}
\usepackage{amsthm}

\usepackage{graphicx}
\usepackage{textcomp}
\usepackage{xcolor}
\usepackage{graphicx}
\usepackage{float}
\usepackage{subfigure}
\usepackage{amsmath}
\usepackage{amsfonts,amssymb}
\usepackage{mathrsfs}
\usepackage{mathtools}
\usepackage{bm}
\usepackage{multirow}
\usepackage{array}
\usepackage{amssymb}
\usepackage{amsmath}
\usepackage{cite}
\usepackage{url}
\usepackage{xcolor}
\usepackage{cite,graphicx,amsmath,amssymb}
\usepackage{subfigure}
\usepackage{fancyhdr}
\usepackage{mdwmath}
\usepackage{mdwtab}
\usepackage{caption}
\usepackage{amsthm}
\usepackage{setspace}
\usepackage{bm}
\usepackage{mathtools}
\usepackage{dsfont}
\usepackage{bbm}
\newtheorem{remark}{Remark}
\newtheorem{theorem}{Theorem}

\newtheorem{lemma}{Lemma}

\newtheorem{corollary}{Corollary}

\makeatletter
\newcommand{\biggg}{\bBigg@{3}}
\newcommand{\Biggg}{\bBigg@{3.5}}
\makeatother
\def\BibTeX{{\rm B\kern-.05em{\sc i\kern-.025em b}\kern-.08em
    T\kern-.1667em\lower.7ex\hbox{E}\kern-.125emX}}
\begin{document}

\title{On the Ergodic Capacity of Reconfigurable Intelligent Surface (RIS)-Aided MIMO Channels}

\author{\IEEEauthorblockN{Chongjun Ouyang$^{*\dag}$, Hao Xu$^{*}$, Xujie Zang$^{*}$, and Hongwen Yang$^{*}$}
$^{*}$Beijing University of Posts and Telecommunications, Beijing, 100876, China\\
$^{\dag}$China Telecom Research Institute, Beijing, 102209, China\\
Email: \{DragonAim, Xu\_Hao, zangxj, yanghong\}@bupt.edu.cn}

\maketitle

\begin{abstract}
Reconfigurable intelligent surfaces (RISs) have emerged as a promising technique to enhance the system spectral efficiency. This paper investigates the ergodic channel capacity (ECC) of an RIS-aided multiple-input multiple-output channel under the assumption that the transmitter-RIS, RIS-receiver, and transmitter-receiver channels contain deterministic line-of-sight paths. Novel expressions are derived to characterize the upper and lower bounds of the ECC. To unveil more system insights, asymptotic analyses are performed to the system ECC in the limit of large signal-to-noise ratio (SNR) and number of reflecting elements (REs). Theoretical analyses suggest that the RIS's deployment can shape the ECC curve by influencing its high-SNR power offset and the ECC can get improved by increasing the number of REs.
\end{abstract}


\section{Introduction}\label{section1}
Reconfigurable intelligent surfaces (RISs) can enhance the system spectral efficiency (SE) with low energy consumption and deployment costs, which is believed to be a promising technology in next-generation networks \cite{Zhang2020}. Specifically, by properly programming its adopted phase shifts, an RIS can control the wireless physical layer channel intelligently and then give it a performance boost \cite{Wu2020}. One of the key scenarios where RIS shines, is for SE improvements in multi-antenna systems \cite{Liu2021}.

Recently, there has been an increasing interest in performing theoretical analyses on the ergodic SE (ESE) of RIS-aided multi-antenna systems \cite{Yu2020,Han2019,Zhi2021,Zhang2021,Wang2021}. For example, a closed-form lower bound was derived to characterize the ESE of an RIS-aided multiuser multiple-input single-output (MISO) channel under Rayleigh fading \cite{Yu2020}. Yet, the RIS is usually deployed on the facade of a high building in the vicinity of transceivers, which yields deterministic line-of-sight (LoS) paths in the RIS-related channels. Thus, it makes more sense to characterize the ESE of RIS-aided systems under fading models that consider the influence of deterministic LoS paths, such as Rician fading. Particularly, a tight upper bound of the ESE of a single-user RIS-aided MISO Rician channel was formulated in \cite{Han2019}, which was further extended to the multiuser case \cite{Zhi2021}. Moreover, the ergodic channel capacity (ECC) of an RIS-aided multiple-input multiple-output (MIMO) Rician channel was investigated in \cite{Wang2021} under the assumption of single-stream transmissions. In fact, the MIMO channel therein can be considered equivalent to a MISO one. As a supplement of this work, a large-system approximation of ECC of an RIS-aided MIMO Rician channel under multi-stream transmission was derived by the replica method \cite{Zhang2021}. Unfortunately, the ECC's calculation therein requires the solutions of a series of fixed-point equations and it is challenging to unveil more insights on the effect of RIS from the analytical results in \cite{Zhang2021}.

To portray the effect of RIS on the ECC and thus fill the aforementioned research gap, this paper evaluates the SE performance of an RIS-aided MIMO Rician channel under multi-stream transmission by formulating upper and lower bounds of the ECC. For gleaning further insights, we perform asymptotic analyses to the ECC in the limit of large signal-to-noise ratio (SNR) and number of reflecting elements (REs). We observe that the RIS can influence the asymptotic ECC through its high-SNR power offset. Moreover, it is also found that increasing the number of REs can reduce the high-SNR power offset and thus enhance the ECC.

\section{System Model}\label{section2}
In a MIMO channel, the transmitter sends signals to a receiver with the aid of an RIS. Assume that the transmitter and receiver are equipped with uniform linear arrays (ULAs) contanining $N_{\text t}$ and $N_{\text r}$ antennas, respectively. Besides, the RIS is equipped with a uniform planar array (UPA) containing $M$ reflecting elements with $M_{\text h}$ ones per row and $M_{\text v}$ ones per column. The signal vector at the receiver end is given by
{\setlength\abovedisplayskip{2pt}
\setlength\belowdisplayskip{2pt}
\begin{align}\label{System_Model}
{\textbf y}=\left(\sqrt{\beta_{\text{t}}\beta_{\text{r}}}{\textbf R}{\bm\Phi}{\textbf T}+\sqrt{\beta_{\text{d}}}{\textbf H}\right){\textbf x}+{\textbf n}={\textbf G}{\textbf x}+{\textbf n},
\end{align}
}where ${\textbf x}\in{\mathbb C}^{N_{\text t}\times 1}$ is the transmitted signal vector satisfying ${\mathbbmss E}\left\{{\textbf x}^{\dag}{\textbf x}\right\}=P$ with $P$ being the total transmit power; ${\textbf T}\in{\mathbb C}^{M\times N_{\text t}}$, ${\textbf R}\in{\mathbb C}^{N_{\text r}\times M}$, and ${\textbf H}\in{\mathbb C}^{N_{\text r}\times N_{\text t}}$ denote the channel matrices from the transmitter to the RIS, from the RIS to the receiver, and from the transmitter to the receiver, respectively; ${\bm\Phi}={\text{diag}}\left\{{\text e}^{\textsf{j}\theta_1},{\text e}^{\textsf{j}\theta_2},\cdots,{\text e}^{\textsf{j}\theta_M}\right\}\in{\mathbb C}^{M\times M}$ denotes the diagonal reflection matrix of the RIS with ${\theta_i}\in\left(-\pi,\pi\right]$, $\forall i=1,2,\cdots,M$; ${\textbf n}\sim{\mathcal{CN}}\left({\textbf 0},\sigma^2{\textbf I}_{N_{\text r}}\right)$ represents the additive white Gaussian noise (AWGN) with $\sigma^2$ being the noise power at each receive antenna; $\beta_{\text{t}}$, $\beta_{\text{r}}$, $\beta_{\text{d}}$ model both path loss and shadowing in the transmitter-RIS, RIS-receiver, and transmitter-receiver channels, respectively; ${\textbf G}=\sqrt{\beta_{\text{t}}\beta_{\text{r}}}{\textbf R}{\bm\Phi}{\textbf T}+\sqrt{\beta_{\text{d}}}{\textbf H}\in{\mathbb C}^{N_{\text r}\times N_{\text t}}$ denotes the effective channel matrix from the transmitter to the receiver.

As stated before, there are deterministic LoS paths in the RIS-related channels in general. To highlight their influence and to make the following analyses more amenable, assume that $\textbf T$ is a pure LoS channel \cite{Nadeem2020}. Moreover, the uncorrelated frequency-flat Rician fading is used to model the propagation environment of $\textbf R$ and $\textbf H$. We comment that the following discussions can be directly extended to the scenario when $\textbf R$ is a LoS channel while $\textbf T$ is a Rician one. By \cite{Paulraj2003}, we have
{\setlength\abovedisplayskip{2pt}
\setlength\belowdisplayskip{2pt}
\begin{align}
&{\textbf T}=\sum\nolimits_{i=1}^{P_{\textbf T}}\frac{t_i}{\sqrt{P_{\textbf T}}}{\textbf a}_{M_{\text h},M_{\text v}}\left(\varphi_{{\text a}_{1,i}},\theta_{{\text a}_{1,i}}\right){\textbf a}_{N_{\text t}}\left(\theta_{{\text d}_{1,i}}\right)^\intercal,\label{Basic_Channel_Matrix}
\end{align}
}where $P_{\textbf T}$ is the number of LoS paths; $\frac{t_i}{\sqrt{P_{\textbf T}}}\in{\mathbb C}$ is the complex gain of the $i$-th path; $t_i$ follow the independent and identically distributed (i.i.d.) complex Gaussian distribution ${\mathcal{CN}}\left(0,1\right)$;
{\setlength\abovedisplayskip{2pt}
\setlength\belowdisplayskip{2pt}
\begin{equation}
\begin{split}
&{\textbf a}_{M_{\text h},M_{\text v}}\left(\varphi,\theta\right)^\intercal=
\left[1,{\text e}^{\textsf{j}2\pi\frac{d_{\text v}}{\lambda}\sin{\theta}},\cdots,{\text e}^{\textsf{j}2\pi\left(M_{\text{v}}-1\right)\frac{d_{\text v}}{\lambda}\sin{\theta}}\right]\\
&\otimes\left[1,{\text e}^{\textsf{j}2\pi\frac{d_{\text h}}{\lambda}\cos\left(\theta\right)\sin\left(\varphi\right)},\cdots,{\text e}^{\textsf{j}2\pi\left(M_{\text{h}}-1\right)\frac{d_{\text h}}{\lambda}\cos\left(\theta\right)\sin\left(\varphi\right)}\right],
\end{split}
\end{equation}
}is the array response of the RIS with $\otimes$ denoting the Kronecker product of matrices, $\lambda$ denoting the wavelength of the carrier frequency, $d_{\text h}$ and $d_{\text v}$ denoting the inner-element space in each row and column of the RIS, respectively; ${\textbf a}_{N_{\text t}}\left(\theta\right)=\left[1,{\text e}^{\textsf{j}2\pi\frac{d_{\text t}}{\lambda}\sin{\theta}},\cdots,{\text e}^{\textsf{j}2\pi\left(N_{\text t}-1\right)\frac{d_{\text t}}{\lambda}\sin{\theta}}\right]^{\intercal}$ is the array response of the ULA at the transmitter with $d_{\text t}$ denoting the inner-element space; $\theta_{{\text a}_{i,i}}$ and $\varphi_{{\text a}_{i,i}}$ are the elevation and azimuth angles of arrival (AoAs), respectively; $\theta_{{\text d}_{i,i}}$ are the angles of departure (AoDs). As for $\textbf R$ and $\textbf H$, we have ${\textbf A}=\sqrt{L_{\textbf A}}\bar{\textbf A}+\sqrt{N_{\textbf A}}\tilde{\textbf A}$, ${\textbf A}\in\left\{{\textbf R},{\textbf H}\right\}$, where $L_{\textbf A}+N_{\textbf A}=1$; $\tilde{\textbf A}$ is the non-LoS (NLoS) component and the elements of $\tilde{\textbf A}$ are i.i.d. complex Gaussian random variables, each with zero mean and unit variance. Like \eqref{Basic_Channel_Matrix}, $\bar{\textbf R}$ and $\bar{\textbf H}$ can be written, respectively, as
{\setlength\abovedisplayskip{2pt}
\setlength\belowdisplayskip{2pt}
\begin{align}
&\bar{\textbf R}=\sum\nolimits_{i=1}^{P_{\textbf R}}\frac{r_i}{\sqrt{P_{\textbf R}}} {\textbf a}_{N_{\text r}}\left(\theta_{{\text a}_{2,i}}\right){\textbf a}_{M_{\text h},M_{\text v}}\left(\varphi_{{\text d}_{2,i}},\theta_{{\text d}_{2,i}}\right)^\intercal,\label{R_LOS}\\
&\bar{\textbf H}=\sum\nolimits_{i=1}^{P_{\textbf H}}\frac{h_i}{\sqrt{P_{\textbf H}}} {\textbf a}_{N_{\text r}}\left(\theta_{{\text a}_{3,i}}\right){\textbf a}_{N_{\text t}}\left(\theta_{{\text d}_{3,i}}\right)^\intercal\label{H_LOS},
\end{align}
}where ${\textbf a}_{N_{\text r}}\left(\theta\right)=\left[1,{\text e}^{\textsf{j}2\pi\frac{d_{\text r}}{\lambda}\sin{\theta}},\cdots,{\text e}^{\textsf{j}2\pi\left(N_{\text r}-1\right)\frac{d_{\text r}}{\lambda}\sin{\theta}}\right]^{\intercal}$ is the array response of the ULA at the receiver with $d_{\text r}$ denoting the inner-element space; $P_{\textbf R}$ and $P_{\textbf H}$ denote numbers of LoS paths; ${r_i}\sim{\mathcal{CN}}\left(0,1\right)$ and ${h_i}\sim{\mathcal{CN}}\left(0,1\right)$; $\theta_{{\text a}_{2,i}}$ and $\theta_{{\text a}_{3,i}}$ denote AoAs; $\varphi_{{\text d}_{2,i}}$, $\theta_{{\text d}_{2,i}}$, and $\theta_{{\text d}_{3,i}}$ denote AoDs.

\section{Ergodic Channel Capacity}
Assume that the instantaneous channel state information (CSI) of ${\textbf G}$ is unknown at the transmitter and perfectly known at the receiver. We comment that this assumption has been widely used in the MIMO literature, especially for the frequency division duplex (FDD) systems. Then, the ECC can be written as
{\setlength\abovedisplayskip{2pt}
\setlength\belowdisplayskip{2pt}
\begin{align}\label{UTk_Ergodic_Rate}
C={\mathbbmss E}\left\{\log_2\det\left({{\textbf I}_{N_{\text r}}+\bar\rho{\textbf G}{\textbf G}^{\dag}}\right)\right\},
\end{align}
}where $\bar\rho=\frac{\rho}{N_{\text t}}$ with $\rho=\frac{P}{\sigma^2}$ being the SNR and the expectation is taken with respect to $\textbf G$. Note that the phase shifts used at the RIS should be computed by either transmitter or receiver, because the RIS has no computation power. Also, it is challenging to estimate $\textbf R$ or $\textbf T$, since the RIS cannot send or detect pilot signals. Recalling the previous assumption, it is clear that neither the transmitter nor receiver has access to the instantaneous CSI of $\textbf R$ and $\textbf T$. As a result, if no more other information of $\textbf R$ and $\textbf T$ is available at the transceivers, random phase shifts are the only choices. In contrast to this, if the statistical CSI of $\textbf H$, $\textbf T$, and $\textbf R$ is available at the transmitter or receiver, then the phase shifts may be designed more properly by utilizing this information. Yet, in both these two cases, we can treat $\theta_i$ ($\forall i$) as some constants when analyzing the ECC. We now intend to evaluate the ECC of the considered channel.

\subsection{New Channel Statistics}
\vspace{-5pt}
\begin{lemma}\label{Lemma1}
The effective channel, $\textbf G$, satisfies
{\setlength\abovedisplayskip{2pt}
\setlength\belowdisplayskip{2pt}
\begin{align}\label{Matrix_Distribute1}
{\text{vec}}\left({\textbf G}^{\dag}\right)\sim{\mathcal{CN}}\left({\text{vec}}\left(\bar{\textbf G}^{\dag}\right),{\textbf I}_{N_{\text r}}\otimes{\bm\Psi}\right),
\end{align}
}where ${\text{vec}}\left(\cdot\right)$ stacks the columns of a matrix into a single vector, $\bar{\textbf G}=\sqrt{\beta_{\text t}\beta_{\text r}L_{\textbf R}}\bar{\textbf R}{\bm\Phi}{\textbf T}+\sqrt{\beta_{\text d}L_{\textbf H}}\bar{\textbf H}\in{\mathbb C}^{N_{\text r}\times N_{\text t}}$, $\bm\Psi={\beta_{\text t}\beta_{\text r}N_{\textbf R}}{\textbf T}^{\dag}{\textbf T}+{\beta_{\text d}N_{\textbf H}}{\textbf I}_{N_{\text t}}\in{\mathbb C}^{N_{\text t}\times N_{\text t}}$, and ${\bm\Psi}\succ0$.
\end{lemma}
\vspace{-5pt}
\begin{IEEEproof}
This lemma can be proven by first stacking $\textbf G$ into a single vector and then performing some basic mathematical manipulations.
\end{IEEEproof}
It follows from \eqref{Matrix_Distribute1} that $\textbf G$ can be written as ${\textbf G}=\bar{\textbf G}+{\textbf X}{\bm\Psi}^{1/2}$ \cite{Gupta2000}, where ${\textbf X}\in{\mathbb C}^{N_{\text r}\times N_{\text t}}$ contains independent ${\mathcal{CN}}\left(0,1\right)$ elements. This further leads to the following remark.
\vspace{-5pt}
\begin{remark}
The effective channel of the considered RIS-aided MIMO system is equivalent to a transmit correlated Rician MIMO channel with its mean matrix giving by $\bar{\textbf G}$ and its transmit correlation matrix giving by $\bm\Psi$.
\end{remark}
\vspace{-5pt}
Generally, it is challenging to derive a concise as well as compact expression to characterize the ECC of a MIMO fading channel. As a compromise, we intend to derive some capacity bounds to evaluate the SE performance of the considered RIS-aided system and use them to unveil more system insights.

\subsection{Capacity Upper Bound}\label{section3a}
Let us introduce some key preliminary results that will be useful in constructing the ECC's upper bound.
\vspace{-5pt}
\begin{lemma}\label{Lemma2}
For ${\textbf A}\in{\mathbb C}^{n\times n}$, let ${\textbf A}_{\mathcal F}$ denote the sub-matrix formed from ${\textbf A}$ by removing columns and rows not indexed by the elements of ${\mathcal F}\subseteq\left\{1,2,\cdots,n\right\}$. Then, we have
{\setlength\abovedisplayskip{2pt}
\setlength\belowdisplayskip{2pt}
\begin{align}\label{Det_Equation_Principle}
\det\left({\textbf I}_n+\lambda{\textbf A}\right)=\sum_{t=0}^{n}\lambda^t\sum_{{\mathcal F}\subseteq\left\{1,2,\cdots,n\right\},\left|{\mathcal F}\right|=t}
\det\left({\textbf A}_{{\mathcal F}}\right),
\end{align}
}where ${\mathcal F}$ is an ordered subset of $\left\{1,\cdots,n\right\}$ with $\left|{\mathcal F}\right|=t$ and $\lambda$ is a constant.
\end{lemma}
\vspace{-5pt}
\begin{IEEEproof}
Please refer to \cite{Aitken1956} for more details.
\end{IEEEproof}
\vspace{-5pt}
\begin{lemma}\label{Lemma3}
Given a random matrix ${\textbf B}\in{\mathbb C}^{p\times q}$ ($p\leq q$) satisfying ${\text{vec}}\left({\textbf B}^{\dag}\right)\sim{\mathcal{CN}}\left({\text{vec}}\left(\bar{\textbf B}^{\dag}\right),{\bm\Omega}\otimes{\textbf I}_{q}\right)$ with ${\mathbbmss E}\left\{\textbf B\right\}=\bar{\textbf B}$ and $\bm\Omega\in{\mathbb C}^{p\times p}$ being a positive definite matrix, then we have
{\setlength\abovedisplayskip{2pt}
\setlength\belowdisplayskip{2pt}
\begin{align}
{\mathbbmss E}\left\{\det\left({\textbf B}{\textbf B}^{\dag}\right)\right\}=\frac{\left(q-L\right)!}{\left(q-p\right)!}
\frac{{\mathsf J}\left(\theta_1,\cdots,\theta_L\right)}{\det\left({\bm\Omega}^{-1}\right)},\label{Wishart_Eq1}
\end{align}
}where ${\mathsf J}\left(\theta_1,\cdots,\theta_L\right)=\det\left({\bm\Delta}\left(\theta_1,\cdots,\theta_L\right)\right)/V\left(\theta_1,\cdots,\theta_L\right)$
$\left\{\theta_i\right\}_{i=1}^{L}$ are the non-zero eigenvalues of ${\bm\Omega}^{-1}\bar{\textbf B}\bar{\textbf B}^{\dag}$, $V\left(\theta_1,\cdots,\theta_L\right)=\prod_{i<j\leq L}\left(\theta_j-\theta_i\right)$, and the $(i,j)$-th element of ${\bm\Delta}\left(\theta_1,\cdots,\theta_L\right)\in{\mathbb C}^{L\times L}$ is given by $\left(q-L+j+\theta_i\right)\theta_i^{j-1}$.
\end{lemma}
\vspace{-5pt}
\begin{IEEEproof}
Please refer to \cite{McKay2005} for more details.
\end{IEEEproof}
\vspace{-5pt}
\begin{lemma}\label{Lemma4}
Given ${\textbf C}\in{\mathbb C}^{p\times q}$ ($p>q$), $\det{\left({\textbf C}{\textbf C}^{\dag}\right)}=0$.
\end{lemma}
\vspace{-5pt}
\begin{IEEEproof}
By using the fact that ${\rm{rank}}\left({\textbf C}{\textbf C}^{\dag}\right)\leq q<p$, this lemma can be found.
\end{IEEEproof}

\subsubsection{$N_{\text r}\geq N_{\text t}$}
When $N_{\text r}\geq N_{\text t}$, we can rewrite \eqref{UTk_Ergodic_Rate} as
{\setlength\abovedisplayskip{2pt}
\setlength\belowdisplayskip{2pt}
\begin{align}
C&={\mathbbmss E}\left\{\log_2\det\left({{\textbf I}_{N_{\text t}}+\bar\rho{\textbf G}^{\dag}{\textbf G}}\right)\right\}\label{UpperBound1_Eq0}\\
&\leq\log_2{{\mathbbmss E}\left\{\det\left({{\textbf I}_{N_{\text t}}+\bar\rho{\textbf G}^{\dag}{\textbf G}}\right)\right\}},\label{UpperBound1_Eq1}
\end{align}
}where \eqref{UpperBound1_Eq0} is based on the Sylvester’s determinant identity and \eqref{UpperBound1_Eq1} is based on the concavity of logarithmic functions. Denoting ${\bm\Sigma}=\bar{\textbf G}^{\dag}\bar{\textbf G}$, the following theorem can be found.
\vspace{-5pt}
\begin{theorem}\label{theorem1}
When $N_{\text r}\geq N_{\text t}$, the ECC of the RIS-aided MIMO channel is upper bounded by
{\setlength\abovedisplayskip{2pt}
\setlength\belowdisplayskip{2pt}
\begin{align}
C&\leq\log_2\left(\sum\nolimits_{t=0}^{N_{\text t}}\bar\rho^{t}\sum\nolimits_{{\mathcal Q}\subseteq\left\{1,\cdots,N_{\text t}\right\},\left|{\mathcal Q}\right|=t}
\det\left({\bm\Psi}_{{\mathcal Q}}\right)\right.\nonumber\\
&\times\left.{\mathsf J}\left(\tilde\theta_1,\cdots,\tilde\theta_{L_{{\mathcal Q}}}\right){\left(N_{\text r}-L_{{\mathcal Q}}\right)!}/{\left(N_{\text r}-t\right)!}\right)\triangleq C_{1},\label{UpperBound1_Eq2}
\end{align}
}where ${\mathcal Q}$ is an ordered subset of $\left\{1,\cdots,N_{\text t}\right\}$ with $\left|{\mathcal Q}\right|=t$ and $\left\{\tilde\theta_i\right\}_{i=1}^{L_{{\mathcal Q}}}$ are the non-zero eigenvalues of ${\bm\Psi}_{{\mathcal Q}}^{-1}{\bm\Sigma}_{{\mathcal Q}}$.
\end{theorem}
\vspace{-5pt}
\begin{IEEEproof}
This theorem can be proved by firstly expanding the expectation in \eqref{UpperBound1_Eq1} based on \eqref{Det_Equation_Principle} and then calculating the resultant expectations based on \eqref{Wishart_Eq1}.
\end{IEEEproof}
\subsubsection{$N_{\text t}> N_{\text r}$}
When $N_{\text t}> N_{\text r}$, it can be deduced from Lemma \ref{Lemma4} that the determinants of some ${\bm\Sigma}={\textbf G}^{\dag}{\textbf G}$ higher order sub-matrices equal zero, which further yields
{\setlength\abovedisplayskip{2pt}
\setlength\belowdisplayskip{2pt}
\begin{align}\label{UpperBound1_Eq3}
\det\left({{\textbf I}_{N_{\text t}}+\bar\rho{\bm\Sigma}}\right)
=\sum\nolimits_{t=0}^{N_{\text r}}\bar\rho^t\sum\nolimits_{{\mathcal Q}_i}
\det\left({\bm\Sigma}_{{\mathcal Q}_i}\right).
\end{align}
}Here, ${\mathcal Q}$ is an ordered subset of $\left\{1,\cdots,N_{\text t}\right\}$ with $\left|{\mathcal Q}\right|=t$. Leveraging \eqref{Wishart_Eq1}, \eqref{UpperBound1_Eq1}, and \eqref{UpperBound1_Eq3}, the capacity upper bound for $N_{\text t}> N_{\text r}$ can be calculated and the main result is summarized in the following theorem.
\vspace{-5pt}
\begin{theorem}\label{theorem2}
When $N_{\text t}> N_{\text r}$, the ECC of the RIS-aided MIMO channel is upper bounded by
{\setlength\abovedisplayskip{2pt}
\setlength\belowdisplayskip{2pt}
\begin{equation}\label{UpperBound2}
\begin{split}
C&\leq\log_2\left(\sum\nolimits_{t=0}^{N_{\text r}}\bar\rho^{t}\sum\nolimits_{{\mathcal Q}\subseteq\left\{1,\cdots,N_{\text t}\right\},\left|{\mathcal Q}_i\right|=t}
\det\left({\bm\Psi}_{{\mathcal Q}}\right)\right.\\
&\times\left.{\mathsf J}\left(\tilde\theta_1,\cdots,\tilde\theta_{L_{{\mathcal Q}}}\right){\left(N_{\text r}-L_{{\mathcal Q}}\right)!}/{\left(N_{\text r}-t\right)!}\right)\triangleq C_{2}.
\end{split}
\end{equation}
}\end{theorem}
\vspace{-5pt}
\subsection{Capacity Lower Bound}
Turning now the capacity lower bound, we establish the following preliminary results before the derivations,.
\vspace{-5pt}
\begin{lemma}\label{Lemma5}
Given a random matrix ${\textbf B}\in{\mathbb C}^{p\times q}$ ($p\leq q$) satisfying ${\text{vec}}\left({\textbf B}^{\dag}\right)\sim{\mathcal{CN}}\left({\text{vec}}\left(\bar{\textbf B}^{\dag}\right),{\textbf I}_{p}\otimes{\textbf I}_{q}\right)$ with ${\mathbbmss E}\left\{\textbf B\right\}=\bar{\textbf B}$. Then, we have
{\setlength\abovedisplayskip{2pt}
\setlength\belowdisplayskip{2pt}
\begin{equation}
\begin{split}
{\mathbbmss E}\left\{\ln\det\left({\textbf B}{\textbf B}^{\dag}\right)\right\}=\sum_{k=0}^{p-1}\psi\left(q-k\right)+{\mathsf F}\left(\theta_1,\cdots,\theta_L\right),\label{Wishart_Eq2}
\end{split}
\end{equation}
}where $\psi\left(x\right)=\frac{{\rm d}}{{\rm d}x}\ln{\Gamma\left(x\right)}$ is the Digamma function \cite[eq. (6.461)]{Gradshteyn}, $\Gamma\left(x\right)=\int_{0}^{\infty}t^{x-1}{\text e}^{-t}{\rm d}t$ is the gamma function \cite[eq. (6.1.1)]{Gradshteyn}, $\left\{\theta_i\right\}_{i=1}^{L}$ are the non-zero eigenvalues of $\bar{\textbf B}\bar{\textbf B}^{\dag}$,
{\setlength\abovedisplayskip{2pt}
\setlength\belowdisplayskip{2pt}
\begin{align}
&{\mathsf F}\left(\theta_1,\cdots,\theta_L\right)=\frac{\sum_{i=1}^{L}{\det{\left(\bar{\bm\Delta}_i\left(\theta_1,\cdots,\theta_L\right)\right)}}}
{V\left(\theta_1,\cdots,\theta_L\right)},\\
&\bar{\bm\Delta}_i\left(\theta_1,\cdots,\theta_L\right)\nonumber\\
&=\left[{\textbf g}\left(\theta_1\right),\cdots,{\textbf g}\left(\theta_{i-1}\right),{\textbf h}\left(\theta_i\right),{\textbf g}\left(\theta_{i+1}\right),\cdots,{\textbf g}\left(\theta_L\right)\right],\\
&{\textbf g}\left(x\right)=\left[1,x,x^2,\cdots,x^{L-1}\right]^{\intercal},\\
&{\textbf h}\left(x\right)=\left[h_1\left(x\right),h_2\left(x\right),\cdots,h_L\left(x\right)\right]^{\intercal},
\end{align}
}\begin{align}
h_i\left(x\right)=x^{i-1}\sum\nolimits_{k=0}^{\infty}\frac{1-\exp\left(-x\right)\sum_{n=0}^{k}\frac{x^n}{n!}}{q-L+i+k}.
\end{align}
\end{lemma}
\vspace{-5pt}
\begin{IEEEproof}
Please refer to \cite{McKay2005} for more details.
\end{IEEEproof}
\vspace{-5pt}
\begin{lemma}\label{Lemma6}
Let ${\textbf Z}\in{\mathbb C}^{q\times q}$ be a positive definite matrix whose eigenvalues are $\lambda_1\geq\cdots\geq\lambda_q>0$. Then for arbitrary matrix ${\textbf X}\in{\mathbb C}^{p\times q}$ ($p\leq q$), we have
{\setlength\abovedisplayskip{2pt}
\setlength\belowdisplayskip{2pt}
\begin{equation}
\begin{split}\label{LogDet_Equation_Principle}
\det\left({\textbf X}{\textbf X}^{\dag}\right)\prod_{i=1}^{p}\lambda_{i}\geq\det\left({\textbf X}{\textbf Z}{\textbf X}^{\dag}\right)\geq\det\left({\textbf X}{\textbf X}^{\dag}\right)\prod_{i=1}^{p}\lambda_{q-i+1}.
\end{split}
\end{equation}
}\end{lemma}
\vspace{-5pt}
\begin{IEEEproof}
Please refer to \cite{Marshall1979} for more details.
\end{IEEEproof}
\subsubsection{$N_{\text r}\geq N_{\text t}$}
By applying the Minkowski’s determinant inequality \cite[eq. (7.8.22)]{Horn} to \eqref{UpperBound1_Eq0} and then leveraging the convexity of $\log_2\left(1+a\exp\left(x\right)\right)$ in $x$ for $a>0$, we have
{\setlength\abovedisplayskip{2pt}
\setlength\belowdisplayskip{2pt}
\begin{align}
C&\geq N_{\text t}{\mathbbmss E}\left\{\log_2\left(1+\bar\rho\exp\left({N_{\text t}^{-1}}\ln\det\left({\textbf G}^{\dag}{\textbf G}\right)\right)\right)\right\}\\
&\geq N_{\text t}\log_2\left(1+\bar\rho\exp\left({N_{\text t}^{-1}}{\mathbbmss E}\left\{\ln\det\left({\textbf G}^{\dag}{\textbf G}\right)\right\}\right)\right).\label{Lower_Bound_Eq1}
\end{align}
}As stated before, ${\textbf G}=\bar{\textbf G}+{\textbf X}{\bm\Psi}^{1/2}$, which yields
{\setlength\abovedisplayskip{2pt}
\setlength\belowdisplayskip{2pt}
\begin{align}
\ln\det\left({\textbf G}^{\dag}{\textbf G}\right)=\ln\det\left(\bm\Psi\right)+\ln\det\left(\tilde{\textbf G}\tilde{\textbf G}^{\dag}\right),\label{Lower_Bound_Eq2}
\end{align}
}where ${\text{vec}}\left(\tilde{\textbf G}^{\dag}\right)\sim{\mathcal{CN}}\left({\text{vec}}\left(\bar{\textbf G}{\bm\Psi}^{-1/2}\right),{\textbf I}_{N_{\text t}}\otimes{\textbf I}_{N_{\text r}}\right)$. By substituting \eqref{Lower_Bound_Eq2} into \eqref{Lower_Bound_Eq1} and calculating the resultant expectation by \eqref{Wishart_Eq2}, we arrive at the following theorem.
\vspace{-5pt}
\begin{theorem}\label{theorem3}
When $N_{\text r}\geq N_{\text t}$, the ECC of the RIS-aided MIMO channel is lower bounded by
{\setlength\abovedisplayskip{2pt}
\setlength\belowdisplayskip{2pt}
\begin{equation}\label{Lower_Bound1}
\begin{split}
&C\geq N_{\text t}\log_2\left(1+\bar\rho\exp\left({N_{\text t}^{-1}}\left(\ln\det\left(\bm\Psi\right)\right.\right.\right.\\
&\left.\left.\left.+\sum\nolimits_{k=0}^{N_{\text t}-1}\psi\left(N_{\text r}-k\right)+{\mathsf F}\left(\alpha_1,\cdots,\alpha_{\tilde{L}}\right)\right)\right)\right)\triangleq C_{3},
\end{split}
\end{equation}
}where $\left\{\alpha_i\right\}_{i=1}^{\tilde{L}}$ are the non-zero eigenvalues of ${\bm\Psi}^{-\frac{1}{2}}\bar{\textbf G}^{\dag}\bar{\textbf G}{\bm\Psi}^{-\frac{1}{2}}$.
\end{theorem}

\subsubsection{$N_{\text t}> N_{\text r}$}
By following the approach in obtaining \eqref{Lower_Bound_Eq1}, we can obtain
{\setlength\abovedisplayskip{2pt}
\setlength\belowdisplayskip{2pt}
\begin{equation}
C\geq N_{\text r}\log_2\left(1+\bar\rho\exp\left({N_{\text r}^{-1}}{\mathbbmss E}\left\{\ln\det\left({\textbf G}{\textbf G}^{\dag}\right)\right\}\right)\right).\label{Lower_Bound_Eq3}
\end{equation}
}It follows from ${\textbf G}=\bar{\textbf G}+{\textbf X}{\bm\Psi}^{1/2}$ that $\ln\det\left({\textbf G}{\textbf G}^{\dag}\right)=\ln\det\left(\tilde{\textbf G}^{\dag}{\bm\Psi}\tilde{\textbf G}\right)$, which together with \eqref{LogDet_Equation_Principle}, yields
{\setlength\abovedisplayskip{2pt}
\setlength\belowdisplayskip{2pt}
\begin{equation}
\ln\det\left({\textbf G}{\textbf G}^{\dag}\right)\geq\ln\det\left(\tilde{\textbf G}^{\dag}\tilde{\textbf G}\right)+\sum\nolimits_{i=1}^{N_{\text r}}\ln{\zeta_{N_{\text t}-i+1}},\label{Lower_Bound_Eq4}
\end{equation}
}where $\zeta_1\geq\cdots\geq\zeta_{N_{\text t}}>0$ are the eigenvalues of $\bm\Psi$. Note that ${\text{vec}}\left(\tilde{\textbf G}\right)\sim{\mathcal{CN}}\left({\text{vec}}\left({\bm\Psi}^{-1/2}\bar{\textbf G}^{\dag}\right),{\textbf I}_{N_{\text r}}\otimes{\textbf I}_{N_{\text t}}\right)$. By substituting \eqref{Lower_Bound_Eq4} into \eqref{Lower_Bound_Eq3} and calculating the resultant expectation based on \eqref{Wishart_Eq2}, we arrive at the following theorem.
\vspace{-5pt}
\begin{theorem}\label{theorem4}
When $N_{\text t}> N_{\text r}$, the ECC of the RIS-aided MIMO channel is lower bounded by
{\setlength\abovedisplayskip{2pt}
\setlength\belowdisplayskip{2pt}
\begin{equation}\label{Lower_Bound2}
\begin{split}
&C\geq N_{\text r}\log_2\left(1+\bar\rho\exp\left({N_{\text r}^{-1}}\left(\sum\nolimits_{i=1}^{N_{\text r}}\ln{\zeta_{N_{\text t}-i+1}}\right.\right.\right.\\
&\left.\left.\left.+\sum\nolimits_{k=0}^{N_{\text r}-1}\psi\left(N_{\text t}-k\right)+
{\mathsf F}\left(\tilde\alpha_1,\cdots,\tilde\alpha_{\bar{L}}\right)
\right)\right)\right)\triangleq C_{4},
\end{split}
\end{equation}
}where $\left\{\tilde\alpha_i\right\}_{i=1}^{\bar{L}}$ are the non-zero eigenvalues of $\bar{\textbf G}{\bm\Psi}^{-1}\bar{\textbf G}^{\dag}$.
\end{theorem}
\vspace{-5pt}
\begin{remark}
When the statistical CSI including $\bar{\textbf G}$ and $\bm\Psi$ is available at the transmitter or receiver, Theorems \ref{theorem1}, \ref{theorem2}, \ref{theorem3}, and \ref{theorem4} can provide insightful guidelines for designing the reflecting matrix to improve the ECC of the considered system.
\end{remark}
\vspace{-5pt}
More specifically, since the ECC defined in \eqref{UTk_Ergodic_Rate} lacks any closed-form expressions, it makes sense to solve the problems given by ${\text P}_i:{\bm\Phi}^{\star}=\arg\max\nolimits_{\bm\Phi}\left\{C_i\right\}$ ($i\in\left\{1,2,3,4\right\}$) for the sake of finding a suboptimal $\bm\Phi$ in terms of maximizing the ECC. Generally, problems (${\text P}_i$) can be solved by using the classical genetic algorithm (GA) \cite{Mitchell1998}. Yet, since this paper focuses more on the theoretical analyses of the ECC instead of the design of $\bm\Phi$, we do not provide more discussions.

\section{Asymptotic Analyses}
We now intend to apply some asymptotic analyses to the previously derived bounds to unveil more system insights. Specifically, two limiting cases including $\rho\rightarrow\infty$ and $M\rightarrow\infty$ will be considered.
\subsection{Asymptotic Analyses in the High-SNR Regime}
\subsubsection{$N_{\text r}\geq N_{\text t}$}
It can be derived from \eqref{UpperBound1_Eq0} that
{\setlength\abovedisplayskip{2pt}
\setlength\belowdisplayskip{2pt}
\begin{align}
\lim\nolimits_{\rho\rightarrow\infty}\left(C-N_{\text t}\log_2\bar\rho\right)=\frac{1}{\ln{2}}{\mathbbmss E}\left\{\ln\det\left({\textbf G}^{\dag}{\textbf G}\right)\right\}.\label{Asym_Eq2}
\end{align}
}Interestingly, it can be found that \eqref{Asym_Eq2} can be also derived from \eqref{Lower_Bound_Eq1} by setting $\rho$ as infinity, which suggests that this lower bound is exact in the high-SNR regime. The ECC in the high-SNR regime can be generally approximated as \cite{Lozano2005}
{\setlength\abovedisplayskip{2pt}
\setlength\belowdisplayskip{2pt}
\begin{equation}\label{Asy_ECC}
C\approx{\mathcal S}_{\infty}\left(\log_2{\rho}-{\mathcal L}_{\infty}\right),
\end{equation}
}where 
${\mathcal S}_{\infty}=\lim_{\rho\rightarrow\infty}\frac{C}{\log_2{\rho}}$ denotes the high-SNR slope in bits/s/Hz/(3 dB), and ${\mathcal L}_{\infty}=\lim_{\rho\rightarrow\infty}\left(\log_2{\rho}-\frac{C}{{\mathcal S}_{\infty}}\right)$ denotes the high-SNR power offset in 3 dB units. Based on \eqref{Wishart_Eq2} and \eqref{Asym_Eq2}, we can evaluate $S_{\infty}$ and $L_{\infty}$ in as follows.
\vspace{-5pt}
\begin{theorem}\label{theorem5}
When $N_{\text r}\geq N_{\text t}$, the high-SNR slope and high-SNR power offset of the RIS-aided MIMO channel are, respectively, given by
{\setlength\abovedisplayskip{2pt}
\setlength\belowdisplayskip{2pt}
\begin{align}
{\mathcal S}_{\infty}=&N_{\text t},\label{high_SNR_Slope}\\
{\mathcal L}_{\infty}=&\log_2{N_{\text t}}-\left({N_{\text t}\ln{2}}\right)^{-1}\left(\ln\det\left(\bm\Psi\right)+\right.\nonumber\\
&\left.+\sum\nolimits_{k=0}^{N_{\text t}-1}\psi\left(N_{\text r}-k\right)+{\mathsf F}\left(\alpha_1,\cdots,\alpha_{\tilde{L}}\right)\right).\label{high_SNR_PowerOffset}
\end{align}
}\end{theorem}
\vspace{-5pt}
\begin{remark}
The results shown in \eqref{high_SNR_Slope} suggest that the high-SNR slope of the ECC curve over logarithmic SNR is $N_{\text t}$. In other words, when $N_{\text r}\geq N_{\text t}$, the ECC increases by approximately $N_{\text t}$ bits/s/Hz for each 3 dB increase in SNR.
\end{remark}
\vspace{-5pt}
It is clear that $N_{\text t}$ is independent of $\bm\Phi$ while $\left\{\alpha_i\right\}_{i=1}^{\tilde L}$ are functions of $\bm\Phi$, which leads to the following remark.
\vspace{-5pt}
\begin{remark}\label{Remark_System_Design}
Equations \eqref{high_SNR_Slope} and \eqref{high_SNR_PowerOffset} imply that the RIS's deployment can shape the ECC curve over SNR by controlling its high-SNR power offset. Moreover, the RIS's deployment cannot influence the high-SNR slope.
\end{remark}
\vspace{-5pt}

\subsubsection{$N_{\text t}> N_{\text r}$}
By following the approach in obtaining \eqref{high_SNR_Slope} and \eqref{high_SNR_PowerOffset}, we find that when $N_{\text t}> N_{\text r}$, the high-SNR slope and high-SNR power offset can be written, respectively, as
{\setlength\abovedisplayskip{2pt}
\setlength\belowdisplayskip{2pt}
\begin{align}
{\mathcal S}_{\infty}&=N_{\text r},\label{high_SNR_Slope2}\\
{\mathcal L}_{\infty}&=\log_2{N_{\text t}}-\frac{1}{N_{\text r}\ln{2}}{\mathbbmss E}\left\{\ln{\det{\left(\tilde{\textbf G}^{\dag}{\bm\Psi}\tilde{\textbf G}\right)}}\right\}.\label{high_SNR_PowerOffset2}
\end{align}
}Equation \eqref{high_SNR_Slope2} implies that when $N_{\text t}> N_{\text r}$, the ECC increases by approximately $N_{\text r}$ bits/s/Hz for each 3 dB increase in SNR. Moreover, by leveraging \textbf{Lemma} \ref{Lemma5} and \textbf{Lemma} \ref{Lemma6} to calculate the expectation in \eqref{high_SNR_PowerOffset2}, we can get the following theorem.
\vspace{-5pt}
\begin{theorem}\label{theorem6}
When $N_{\text t}> N_{\text r}$, the high-SNR power offset of the RIS-aided MIMO channel is upper bounded by
{\setlength\abovedisplayskip{2pt}
\setlength\belowdisplayskip{2pt}
\begin{align}\label{high_SNR_PowerOffset2_UB}
{\mathcal L}_{\infty}\leq\log_2{N_{\text t}}-\frac{\sum_{i=1}^{N_{\text r}}\ln{\zeta_{N_{\text t}-i+1}}
+{\mathsf L}\left(\tilde\alpha_1,\cdots,\tilde\alpha_{\bar{L}}\right)}{N_{\text r}\ln{2}}
\end{align}
}and lower bounded by
{\setlength\abovedisplayskip{2pt}
\setlength\belowdisplayskip{2pt}
\begin{align}
{\mathcal L}_{\infty}\geq\log_2{N_{\text t}}-\frac{\sum_{i=1}^{N_{\text r}}\ln{\zeta_{i}}+
{\mathsf L}\left(\tilde\alpha_1,\cdots,\tilde\alpha_{\bar{L}}\right)}{N_{\text r}\ln{2}},
\end{align}
}where ${\mathsf L}\left(\tilde\alpha_1,\cdots,\tilde\alpha_{\bar{L}}\right)=\sum_{k=0}^{N_{\text r}-1}\psi\left(N_{\text t}-k\right)+{\mathsf F}\left(\tilde\alpha_1,\cdots,\tilde\alpha_{\bar{L}}\right)$.
\end{theorem}
\vspace{-10pt}
\subsection{Asymptotic Analyses in the Limit of Many REs}\label{Section_Asym_B}
We first rewrite \eqref{UTk_Ergodic_Rate} as
{\setlength\abovedisplayskip{2pt}
\setlength\belowdisplayskip{2pt}
\begin{align}
C={\mathbbmss E}\left\{\log_2\det\left({{\textbf I}_{N_{\text r}}+\bar\rho M\left(\frac{{\textbf G}}{\sqrt{M}}\right)\left(\frac{{\textbf G}}{\sqrt{M}}\right)^{\dag}}\right)\right\}.
\end{align}
}Since ${\text{vec}}\left({\textbf G}^{\dag}\right)\sim{\mathcal{CN}}\left({\text{vec}}\left(\bar{\textbf G}^{\dag}\right),{\textbf I}_{N_{\text r}}\otimes{\bm\Psi}\right)$, we can obtain ${\text{vec}}\left(\frac{{\textbf G}^{\dag}}{\sqrt{M}}\right)\sim{\mathcal{CN}}\left({\text{vec}}\left(\frac{\bar{\textbf G}^{\dag}}{\sqrt{M}}\right),{\textbf I}_{N_{\text r}}\otimes\frac{{\bm\Psi}}{M}\right)$. Based on \cite[Page 9]{Zhi2021}, $\left|{\textbf a}_{M_{\text h},M_{\text v}}\left(\varphi_{{\text d}_{2,i}},\theta_{{\text d}_{2,i}}\right)^\intercal{\bm\Phi}{\textbf a}_{M_{\text h},M_{\text v}}\left(\varphi_{{\text a}_{1,j}},\theta_{{\text a}_{1,j}}\right)\right|$ ($\forall i,j $) is a bounded value under random phase shifts, which together with \eqref{Basic_Channel_Matrix}, \eqref{R_LOS}, and \eqref{H_LOS}, yields $\lim_{M\rightarrow\infty}\frac{\bar{\textbf G}^{\dag}}{\sqrt{M}}={{\textbf 0}_{N_{\text t}\times N_{\text r}}}$. Besides,
{\setlength\abovedisplayskip{2pt}
\setlength\belowdisplayskip{2pt}
\begin{equation}
\lim_{M\rightarrow\infty}\frac{{\bm\Psi}}{M}={\beta_{\text t}\beta_{\text r}N_{\textbf R}}\sum_{i=1}^{P_{\textbf T}}\frac{\left|t_i\right|^2}{P_{\textbf T}}{\textbf a}_{N_{\text t}}\left(\theta_{{\text d}_{1,i}}\right)^{\ast}{\textbf a}_{N_{\text t}}\left(\theta_{{\text d}_{1,i}}\right)^{\intercal}\triangleq{\bm\Upsilon}.\nonumber
\end{equation}
}Thus, $\lim_{M\rightarrow\infty}{\text{vec}}\left(\frac{{\textbf G}^{\dag}}{\sqrt{M}}\right)\sim{\mathcal{CN}}\left({\textbf 0},{\textbf I}_{N_{\text r}}\otimes{\bm\Upsilon}\right)$. It follows that as $M\rightarrow\infty$, the ECC satisfies $C-\dot{C}\left(\rho\right)\xrightarrow[M\rightarrow\infty]{a.s.}0$, where
{\setlength\abovedisplayskip{2pt}
\setlength\belowdisplayskip{2pt}
\begin{align}\label{Asym_ECC}
\dot{C}\left(\rho\right)={\mathbbmss E}_{\textbf X}\left\{\log_2\det\left({\textbf I}_{N_{\text r}}+M{\rho}/{N_{\text t}}{\textbf X}{\bm\Upsilon}{\textbf X}^{\dag}\right)\right\}
\end{align}
}with ${\textbf X}\in{\mathbb C}^{N_{\text r}\times N_{\text t}}$ containing independent ${\mathcal{CN}}\left(0,1\right)$ elements. We now intend to consider the case where the transmit power is scaled down by a factor of $1/M^{\alpha}$. From \eqref{Asym_ECC}, we have $\lim_{M\rightarrow\infty}\dot{C}\left({\rho}{M^{-\alpha}}\right)=\infty$ for $\alpha<1$, $\lim_{M\rightarrow\infty}\dot{C}\left({\rho}{M^{-\alpha}}\right)=0$ for $\alpha>1$, and $\lim_{M\rightarrow\infty}\dot{C}\left({\rho}/{M}\right)={\mathbbmss E}_{\textbf X}\left\{\log_2\det\left({\textbf I}_{N_{\text r}}+\rho/N_{\text t}{\textbf X}{\bm\Upsilon}{\textbf X}^{\dag}\right)\right\}$.
\vspace{-5pt}
\begin{remark}
It is clear that $\dot{C}\left({\rho}{M^{-\alpha}}\right)$ is a non-zero constant if and only if $\alpha=1$, which suggests that in the large limit of $M$, with no degradation in the ECC, the total SNR or transmit power can be cut down at most ${1}/{M}$.
\end{remark}
\vspace{-5pt}
It is challenging to derive a close-form expression for $\tilde{C}=\dot{C}\left(\rho/M\right)$. Fortunately, by following similar steps outlined in Section \ref{section3a}, a closed-form upper bound of $\tilde{C}$ is available. For the sake of brevity, we omit the details here.
\subsubsection{$N_{\text r}\geq N_{\text t}$}
With the aid of \eqref{UpperBound1_Eq2} and \eqref{Lower_Bound1}, the capacity bounds for $N_{\text r}\geq N_{\text t}$ can be directly obtained. Furthermore, the following corollary can be found.
\vspace{-5pt}
\begin{corollary}\label{corollary1}
In the large limit of $M$, the high-SNR slope and high-SNR power offset for the case of $N_{\text r}\geq N_{\text t}$ are given by ${\mathcal S}_{\infty}=N_{\text t}$ and ${\mathcal L}_{\infty}=\log_2\frac{N_{\text t}}{M}-\frac{1}{N_{\text t}\ln{2}}\left(\ln\det\left(\bm\Upsilon\right)+\sum_{k=0}^{N_{\text t}-1}\psi\left(N_{\text r}-k\right)\right)$, respectively.
\end{corollary}
\vspace{-5pt}
We comment that in obtaining Corollary \ref{corollary1}, we have assumed that $\left|\ln\det\left(\bm\Upsilon\right)\right|<\infty$ for simplicity. The results in Corollary \ref{corollary1} suggest that increasing REs' number can decrease the high-SNR power offset and thus enhance the ECC.
\subsubsection{$N_{\text t}> N_{\text r}$}
By following the approach in the last section, the capacity bounds under the scenario of $N_{\text t}> N_{\text r}$ can be also discussed. Thus, we do not provide further detailed statements.

\section{Numerical Results}
\begin{figure}[!t]
\vspace{-5pt}
    \centering
    \subfigbottomskip=0pt
	\subfigcapskip=-5pt
\setlength{\abovecaptionskip}{0pt}
   \subfigure[$N_{\text r}\geq N_{\text t}$.]
    {
        \includegraphics[height=0.175\textwidth]{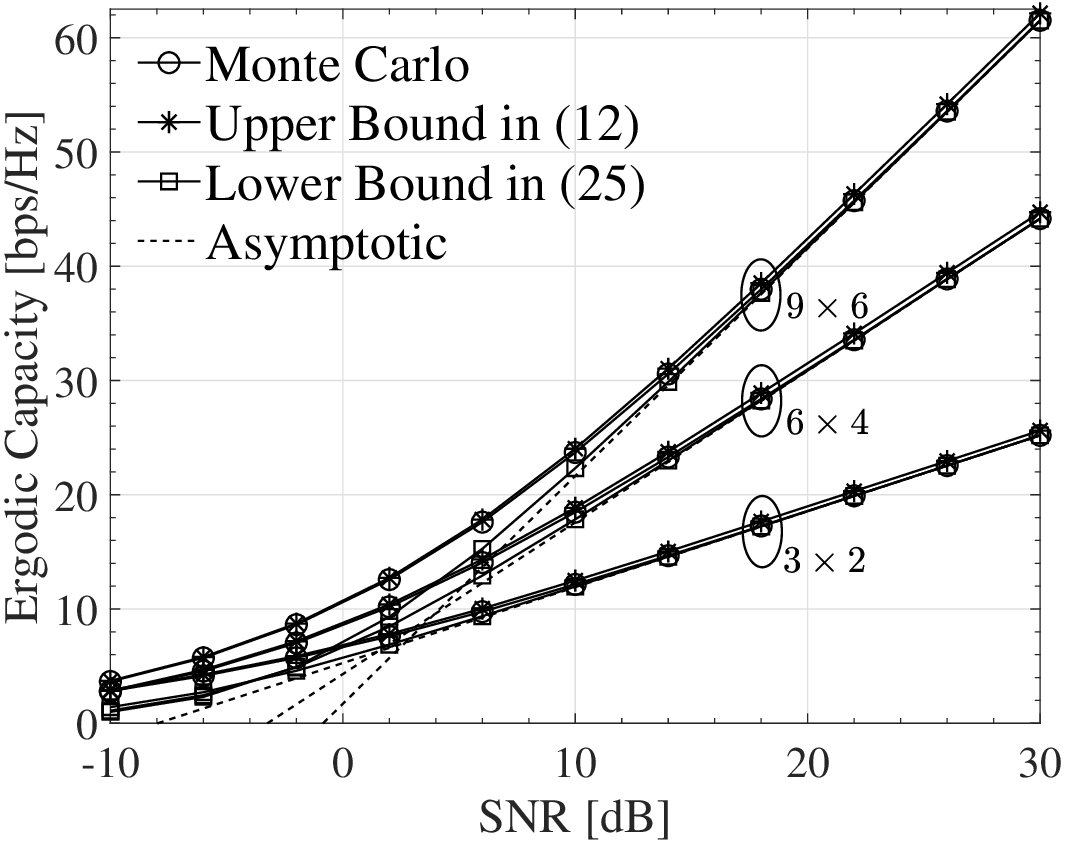}
	   \label{fig1a}	
    }
    \subfigure[$N_{\text t}> N_{\text r}$.]
    {
        \includegraphics[height=0.175\textwidth]{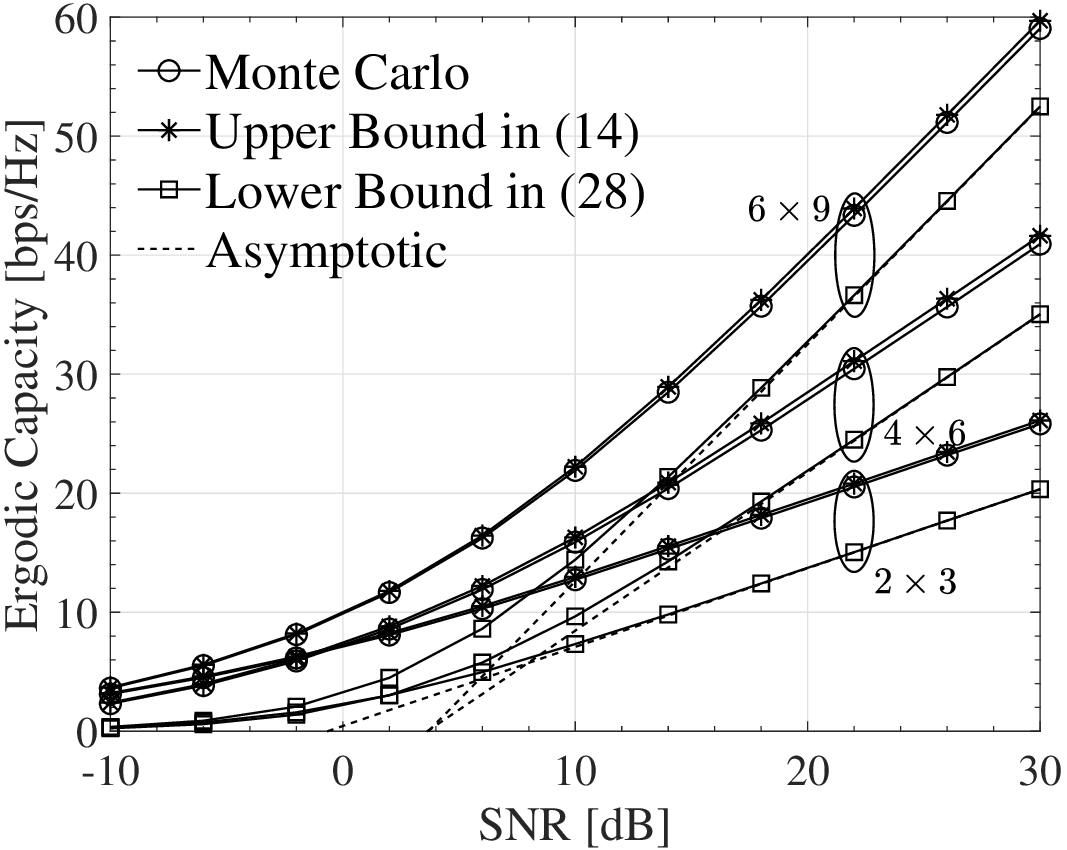}
	   \label{fig1b}	
    }
   \caption{ECC bounds with $M=16$. The asymptotic results in {\figurename} {\ref{fig1a}} are calculated by \eqref{Asy_ECC}, \eqref{high_SNR_Slope}, and \eqref{high_SNR_PowerOffset}, whereas the asymptotic results in {\figurename} {\ref{fig1b}} are calculated by \eqref{Asy_ECC}, \eqref{high_SNR_Slope2}, and \eqref{high_SNR_PowerOffset2_UB}.}
    \label{figure1}
    \vspace{-10pt}
\end{figure}

\begin{figure}[!t]
    \centering
    \subfigbottomskip=0pt
	\subfigcapskip=-5pt
\setlength{\abovecaptionskip}{0pt}
   \subfigure[$\rho=10$ dB.]
    {
        \includegraphics[height=0.18\textwidth]{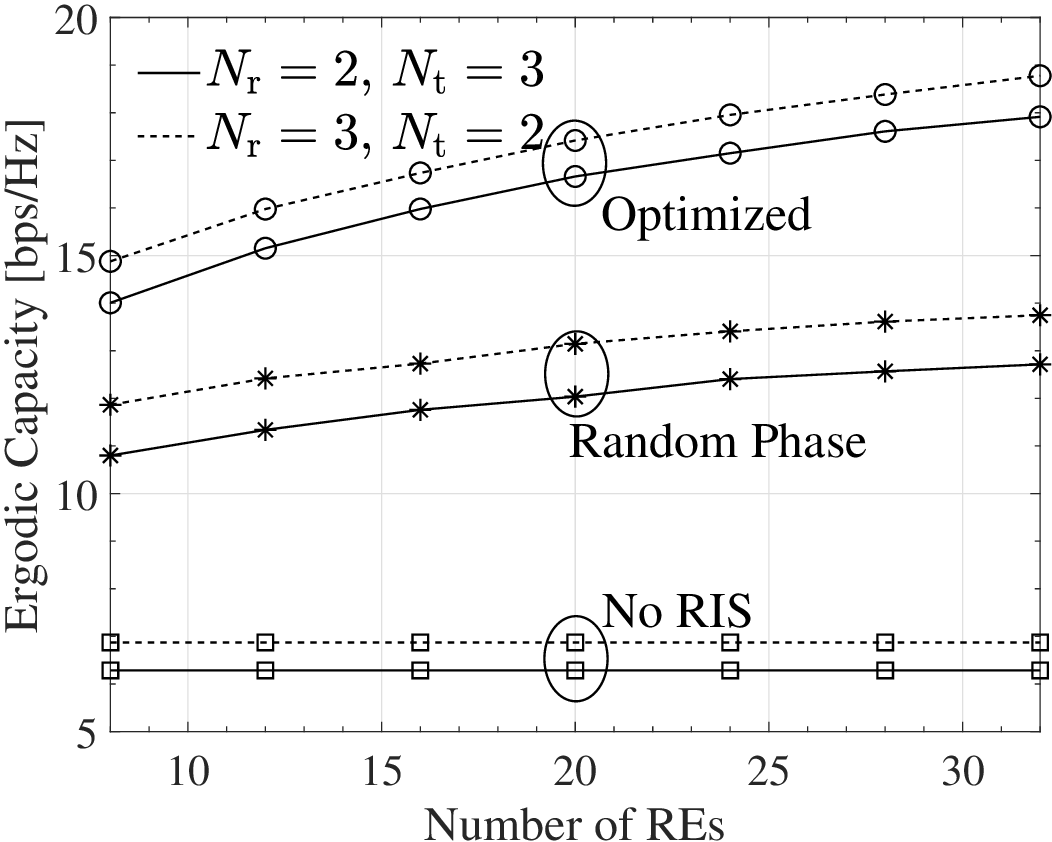}
	   \label{fig2a}	
    }
    \subfigure[$\rho=E/M$ and $E=10$ dB.]
    {
        \includegraphics[height=0.18\textwidth]{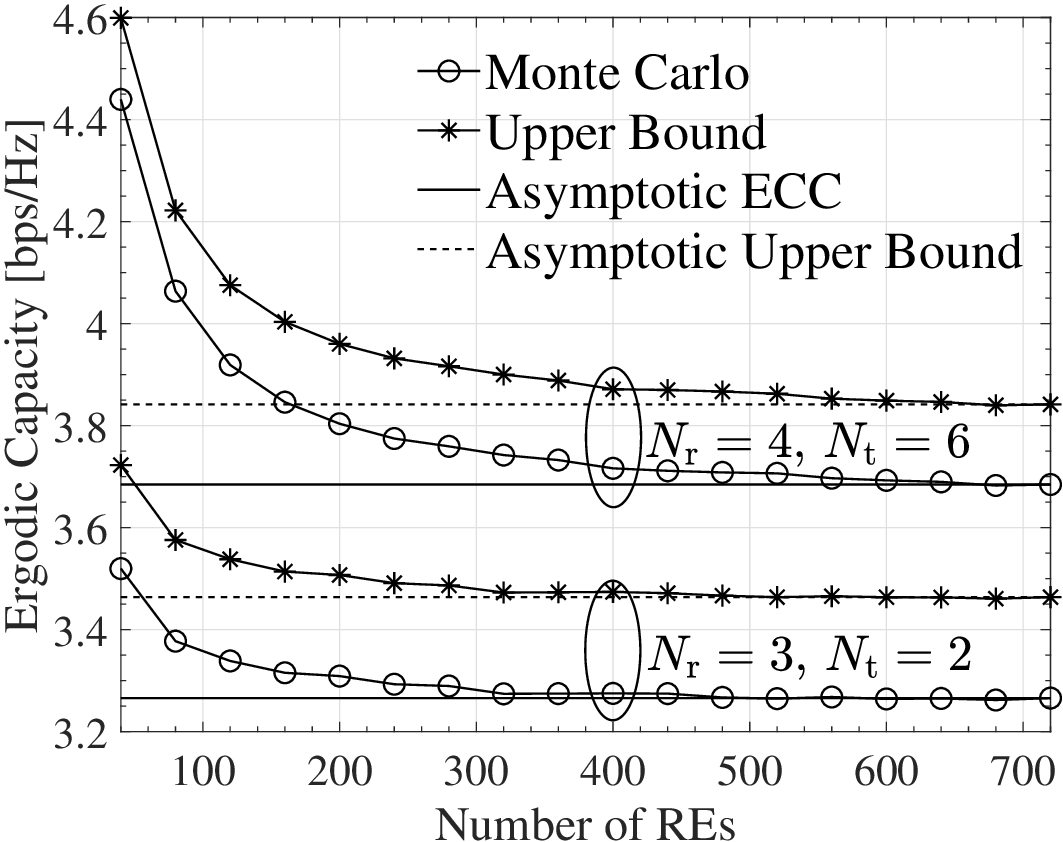}
	   \label{fig2b}	
    }
   \caption{The ECC versus the number of REs with $M_{\text h}=4$. The asymptotic ECC ($\tilde{C}$) in {\figurename} {\ref{fig2b}} is obtained by simulation, whereas the asymptotic upper bound is obtained by following similar steps in deriving \eqref{UpperBound1_Eq2} and \eqref{UpperBound2}.}
    \label{figure2}
    \vspace{-10pt}
\end{figure}

Numerical results are provided in this part to verify the previous analyses. During the simulation, we set $d_{\text t}=d_{\text r}=d_{\text v}=d_{\text h}=\frac{\lambda}{2}$, $\beta_{\text t}=\beta_{\text r}=\beta_{\text d}=1$, $P_{\textbf R}=P_{\textbf T}=1$, $P_{\textbf H}=2$, and $L_{\textbf R}=L_{\textbf H}=\frac{2}{3}$. Moreover, angles of the LoS components in the assistant channel is randomly set within $\left[0,2\pi\right)$.

{\figurename} {\ref{figure1}} plots the ECC of the RIS-aided MIMO systems under random phase shifts for $M=16$ with $M_{\text h}=4$. The case of $N_{\text r}\geq N_{\text t}$ is presented in {\figurename} {\ref{fig1a}}. As can be seen, the derived upper bound is tight for all SNR ranges, which, thus, serves as a good performance metric. Moreover, the lower bound is tight in the high-SNR regime and the asymptotic results can track the simulations exactly in the high-SNR region, which is consistent with the former analyses. {\figurename} {\ref{fig1b}} plots the ECC for $N_{\text r}< N_{\text t}$. It can be seen from this graph that the derived upper bound is still tight for all SNR ranges. Yet, the derived capacity lower bound is relatively looser than the upper one. Taken this two figures together, it can be concluded that the derived upper bound can serve as a tight bound to benchmark the performance of RIS-aided MIMO systems.

Since the derived upper bounds are tight, it makes sense to leverage them to optimize the phase shifts. {\figurename} {\ref{fig2a}} compares the ECC achieved by optimized, random, and no phase shifts in terms of the number of REs with $\rho=10$ dB. Particularly, the phase shifts in {\figurename} {\ref{fig2a}} are optimized by applying the genetic algorithm to the upper bounds ($C_1$ and $C_2$). As shown, the optimized ECC is much higher than the ECC achieved by random phase shifts, which highlights the importance of phase shifts design. Moreover, it can be seen that the ECC can get improved by increasing the number of REs. {\figurename} {\ref{fig2b}} presents the ECC with scaled-down power $\rho={E}/{M}$, where $E=10$ dB. For simplicity, we only provide the results under random phase shifts. As shown, when $M\rightarrow\infty$, the ECC converges to its asymptotic value, namely $\tilde{C}$, which, thus, verifies our previous derivations.

\section{Conclusion}
\label{section6}
Novel expressions were derived to characterize the upper and lower bounds of the ECC of RIS-aided MIMO channels with deterministic LoS paths. To unveil more system insights, asymptotic analyses were performed to the system ECC in the large limit of the SNR and number of REs. It was found that the RIS can shape the ECC curve via influencing its high-SNR power offset and the ECC increases with the number of REs.

\end{document}